\documentstyle[aps,prd,twocolumn,epsf,epsfig]{revtex}

\newcommand \bea{\begin{eqnarray}}
\newcommand \eea{\end{eqnarray}}
\newcommand \ga{\raisebox{-.5ex}{$\stackrel{>}{\sim}$}}
\newcommand \la{\raisebox{-.5ex}{$\stackrel{<}{\sim}$}}
\newcommand{\av}[1]{\langle{#1}\rangle}

\begin{document}
\twocolumn[\hsize\textwidth\columnwidth\hsize
\csname@twocolumnfalse%
\endcsname
\draft
\title{Phases of Bosons or Fermions in confined optical lattices}
\author{H. Heiselberg}
\address{Univ. of S. Denmark, Campusvej 55, DK-5230 Odense M, Denmark}
%\address{Danish Defense Research Establishment, Ryvangsalle' 1, 
%DK-2100 Copenhagen \O, Denmark}
\maketitle
\begin{abstract}
Phases of Bose or Fermi atoms in optical lattices confined in harmonic
traps are studied within the Thomas-Fermi approximation. Critical radii
and particle number for onset of Mott insulator states are calculated
and phase diagrams shown in 1D, and estimated for 2 and 3D. Methods to
observe these and novel phases such as d-wave superconductivity are discussed.
Specifically the collective modes are calculated. 
\\
\vspace{-.5cm}
\end{abstract}
\vskip1pc]

\section{Introduction}

Cold quantum systems in optical lattice have been realized with Bose
and Fermi atoms, and superfluid transitions to Mott insulator (MI)
solids at filling up to five
\cite{Greiner,Stoferle,Gerbier,Thalhammer,Folling,Campbell}. Luttinger
and Luther-Emery liquids and MI solids appear for 1D fermions whereas
in 2D it is believed that d-wave superconductivity (dSC) appears
between metal and antiferromagnetic (AFM) phases
\cite{Scalapino,Hofstetter}, which is responsible for the high
temperature superconductivity.  Likewise quantum spin phases
\cite{Duan}, supersolids \cite{Scarola}, fractional quantum Hall
phases \cite{Sorensen,FQHE}, mixed \cite{Lukin} and other phases may
be studied.

By tuning the barriers of the optical lattices and the
interaction strengths near or far from Feshbach resonances, the Fermi
\cite{Hubbard}
and Bose \cite{Fisher} 
Hubbard models can be realized 
in 1, 2 and 3 dimensions \cite{Jaksch,Zwerger}
for any interaction and hopping strengths. These are the
on-site interaction energy $U$ and the hopping parameter $t$ to nearest site. 
Both can be expressed in terms of the lattice
potential and particle interactions such as the s-wave scattering length
(see, e.g. \cite{Jaksch,Zwerger}). 
The optical lattices are in experiments usually confined in a trap
typically in the form of a harmonic potential $V_i=V_2r_i^2$.  Here, the
lattice point $r_i=id$ is an integer times the lattice spacing $d$.
The optical lattices are then clean systems with controlled disorder.
A number of MI, superfluid and other phases appear as the densities 
increase towards the center of the trap 
\cite{Jaksch,Zwerger,Paredes,Campo,Drummond,Xianlong,Rigol}.

It is the purpose here to calculate the phase diagram for these models
in the Thomas-Fermi approximation (TF) at zero temperature, which can
be done analytically in 1D for strong and weak couplings $U/t$ and
approximated between these limits. The corresponding phases in 2D and
3D will also be discussed. Finally, the 1D collective modes are
calculated for bosons and fermions in harmonically confined optical
lattices.

\section{Equation of states in the Thomas-Fermi limit}

In the limit of many particles in a shallow
confining harmonic potential, the TF (or local density
approximation) applies because the length scales over which the
trapping potential and density varies are long compared to phase
boundaries \cite{PS} and the lattice spacing.
As will be shown below, the size of the system
scales as $R\sim\sqrt{t/V_2}$, and the total number of particles scales as
$N\sim R/d$. Therefore the average filling fraction scales as 
$\sim N/R\simeq Nd\sqrt{V_2/t}$,
and can be held fixed in the TF limit: $N\to\infty$ and
$V_2\to 0$.

In TF the total chemical potential is given by the sum of the
trap potential and the local chemical potential $\mu(\rho)=d\varepsilon/d\rho$,
where $\varepsilon$ and  $\rho(r)$ are the local energy and density 
per site, i.e.
\bea \label{mu}
  \mu_{Tot} = \mu(\rho,U) +V_2r^2 
      = \mu(0,U)+V_2R^2 \,.
\eea
It must be constant over the lattice and can therefore be set to its value
at the edge or radius $R$ of the atomic gas, which gives the last equation in
(\ref{mu}). As we shall see later $\mu(\rho=0,U)=-zt$ for both repulsive 
fermions and bosons,
where $z=2D$ is the number of neighbors in $D$ dimensions.

\subsection{1D Fermions}

We start with a 1D system of spin-balanced fermions with repulsive 
interactions described by the Hubbard model \cite{Hubbard} 
\bea
 H= -t\sum_{\av{ij}\sigma} \hat{a}_{i\sigma}^\dagger \hat{a}_{j\sigma} \,+\,
  U\sum_i \hat{n}_{i\uparrow} \hat{n}_{i\downarrow} \,,
\eea
where, $\hat{a}_{i\sigma}$ is the usual Fermi 
creation operators, $n_{i\sigma}=\hat{a}_{i\sigma}^\dagger \hat{a}_{i\sigma}$ 
the density, and $\av{ij}$ sums over nearest neighbors $j=i\pm1$. 

The ground state energy per site at zero temperature can
be calculated \cite{Lieb} from the Bethe ansatz.
A good approximation for densities $\rho\le1$ and all $U\ge0$ is \cite{Campo}
\bea \label{eF}
 \mu(\rho,U)= -2t\cos(\pi\rho/\beta) \,.
\eea
It is exact in both limits:
$\beta(U\to\infty)=1$ and $\beta(U\to0)=2$.
For densities $1<\rho\le2$ the chemical potential is:
$\mu(\rho,U)=U-\mu(2-\rho,U)$. \cite{Lieb}
The phase is a band insulator (BI) for $\rho=2$, a MI for $\rho=1$ and 
a Luttinger liquid (LL) \cite{Luttinger} otherwise.
The MI phase has a
chemical potential gap $\Delta\mu=U+4t\cos(\pi/\beta)$.
It must match \cite{Lieb}
\bea \label{dmu}
  \Delta\mu(U) = U-4t+8t\int_0^\infty 
       \frac{J_1(x)}{x[1+\exp(x U/2t)]}dx \,,
\eea
where $J_1$ is the 1st Bessel function.
Alternatively, one could  as in \cite{Campo} require that the energy:
$\epsilon(1,U)=-2t(\beta/\pi)\sin(\pi/\beta)$, matches the exact result
of the Bethe ansatz \cite{Lieb}. We prefer matching the chemical potentials
because it insures that the Mott transitions are correct. 
The chemical potential is discontinuous at $\rho=1$ by an amount
\bea
  \Delta\mu=U-4t+8\ln(2)t^2/U+{\cal O}(U^{-2}) \, , 
\eea for $U\gg t$
whereas in the weakly repulsive limit the gap vanishes exponentially as
\bea
  \Delta\mu=\frac{}{\pi}\sqrt{Ut}\exp(-2\pi t/U) \,.
\eea
The density is now found by inserting (\ref{eF}) in (\ref{mu}).
For $\rho\le1$ we obtain
\bea \label{rho}
   \rho(r) = \frac{2\beta}{\pi} \arcsin(\sqrt{R^2-r^2}/R_c) \,,
\eea
where $R_c=2\sqrt{t/V_2}$. 
The gap in the chemical potential Eq. (\ref{mu}) requires a 
density plateau at $\rho=1$ \cite{Campo,Drummond,Xianlong,Rigol}.
We can now relate 
$\rho(0)$ to the radius $R$ and the corresponding total number of atoms 
\bea
   N=\frac{2}{d} \int^R_0 \rho(r)dr 
\eea
in the confined lattice for any $U/t$.

We are particularly interested in the critical radii $R_n^{\pm}$ and the
corresponding critical number of atoms
$N_n^{\pm}=(2/d)\int_0^{R_n^\pm}\rho(r)dr$, 
right when the MI plateaus of filling $n$ first appear (-) 
in the center of the trap and disappear (+) forming a MI shell. 
The MI phase with
$\rho(0)=n=1$ first appear in the center of the trap when $N=N_1^-$ and
persist up to $N=N_1^+$ where after the core is a LL with $\rho>1$ until
$N=N_2^-$ where a BI with $\rho=2$ forms in the core.

For $U\gg t$ ($\beta\to1$) the MI first appear in the center
when $R_1^-=R_c$ according to Eq. (\ref{rho}), and the corresponding
number of particles is $N_1^-=N_c$, where
\bea
   N_c = \frac{4R_c}{\pi d} = \frac{8}{\pi d}\sqrt{\frac{t}{V_2}} \,.
\eea
Likewise, it follows that
for $U\ll t$, where
$\beta=2$ and $\Delta\mu=0$, the critical MI 
radius is shorter $R_1^+=R_c/\sqrt{2}$
whereas the BI radius is 
$R_2=R_c$. The corresponding critical number of atoms are
\bea
  N_1^-=N_1^+=(4R_c/\pi d)\int^1_0\arccos(x^2)dx\simeq 0.847N_c \,,
\eea  
and $N_2^-=2N_c$.
 
Returning to 
 $U\gg t$  we must add the chemical potential gap
$\Delta\mu=U-4t$ to Eq. (\ref{mu}) when $\rho(r)>1$. 
The critical radius, where then density
just starts to exceed unity $\rho=1_+$ in the core is:
$R_{1}^+=R_c\sqrt{U/4t}$, whereas the BI with $\rho=2$ appears when
$R_{2}^-=R_c\sqrt{U/4t+1}$.
The corresponding number of atoms is
\bea 
  N_{1}^+=\frac{2}{d}\sqrt{U/V_2} = N_c\, \frac{\pi}{4}\sqrt{U/t} \,,
\eea 
and 
\bea 
  N_2^-=N_c\,\left[\frac{\pi}{4}\sqrt{U/t}+1\right] \,.
 \eea
We emphasize that the calculated values for the critical fillings
$N_1^-$, $N_1^+$ and $N_2^-$ 
and the corresponding radii are
exact in both the weak and strong coupling limits within TF.

For general $U$ we calculate $\Delta\mu(U)$ from Eq. (\ref{dmu}) to
find $\beta(U)$.  Inserting (\ref{eF}) in (\ref{mu}) then gives the
density, critical radii and the critical fillings as shown in
Fig. 1. They interpolate smoothly between the analytical results
$U\to0$ and $U\gg t$ given above.

The numerical calculations in Refs. \cite{Scalettar,Drummond} find
similar phase diagrams as the TF continuum limit $N\to\infty$ discussed above, 
however, with a
few important differences.  The finite number of particles makes it
impossible to resolve the exponentially small gap for $U\to 0$.
The MI plateau in the density at $\rho=1$ does not extend over a
sufficient number of lattice sites, $R_1^+-R_1^-\la d$, and the phase
boundary cannot be resolved.  Therefore, the critical point where the
$\rho=1$ MI state vanishes moves from $U_1=0$ in TF to $U_1\simeq
0.8t$ in \cite{Drummond} and $U_1\simeq 3t$ in \cite{Scalettar}. As a
consequence the phase diagrams of Refs. \cite{Scalettar,Drummond}
acquire an additional phase transition at $U_1$ and $N\ga N_1^-$ between
phases with and without a ``resolvable'' $\rho=1$ Mott transition.  A
similar phenomenon will occur at finite temperature, where the MI solid
melts and the critical point moves to a finite $U$ since
$\Delta\mu(U)\simeq k_BT$.

The attractive Hubbard model ($U\le0$) tends to pair atoms in a
Luther-Emery liquid (LE) \cite{Luther}.  The trap potential breaks
translational invariance forming an atomic density wave (ADW)
phase with oscillating density: $\rho(r_i)=\bar{\rho}+\delta\rho
\cos(2\pi r_i/\lambda_{ADW})$ \cite{Luther,Campo,Xianlong}. Here,
$\bar{\rho}$ is the average density, $\delta\rho$ is the amplitude and
$\lambda_{ADW}=2d/\bar{\rho}$ the wavelength.  The TF can be applied
for the average density $\bar{\rho}$ as long as $R\gg\lambda_{ADW}$.
As described in \cite{Campo} the parametrization
\bea \label{LE}
  \mu = U/2 -2 \alpha^2 t \cos(\pi\bar{\rho}/2) \,,
\eea
is a good approximation for the EoS at all densities and
$U\le0$. It satisfies the known properties for weak and strong 
attraction \cite{Lieb}, when the prefactor is chosen as:
$\alpha^2(U)=\pi\int_0^\infty J_0(x)J_1(x)/(1+\exp(x|U|/2))(dx/x)$.
It decreases from $\alpha(U=0)=1$ to  
$\alpha(U\to -\infty)=\sqrt{\pi\ln(2)}t/|U|$.
No MI are formed in the LE at $\bar{\rho}=1$ whereas a
BI appears when $\rho=2$ with radius 
$R_2=2\alpha t/V_2=\alpha R_c$.
The corresponding number of particles is
\bea
 N_2^-=2\alpha N_c \,,
\eea
which is also shown in Fig. 1.

\vspace{-0.5cm}
\begin{figure}
\begin{center}
\psfig{file=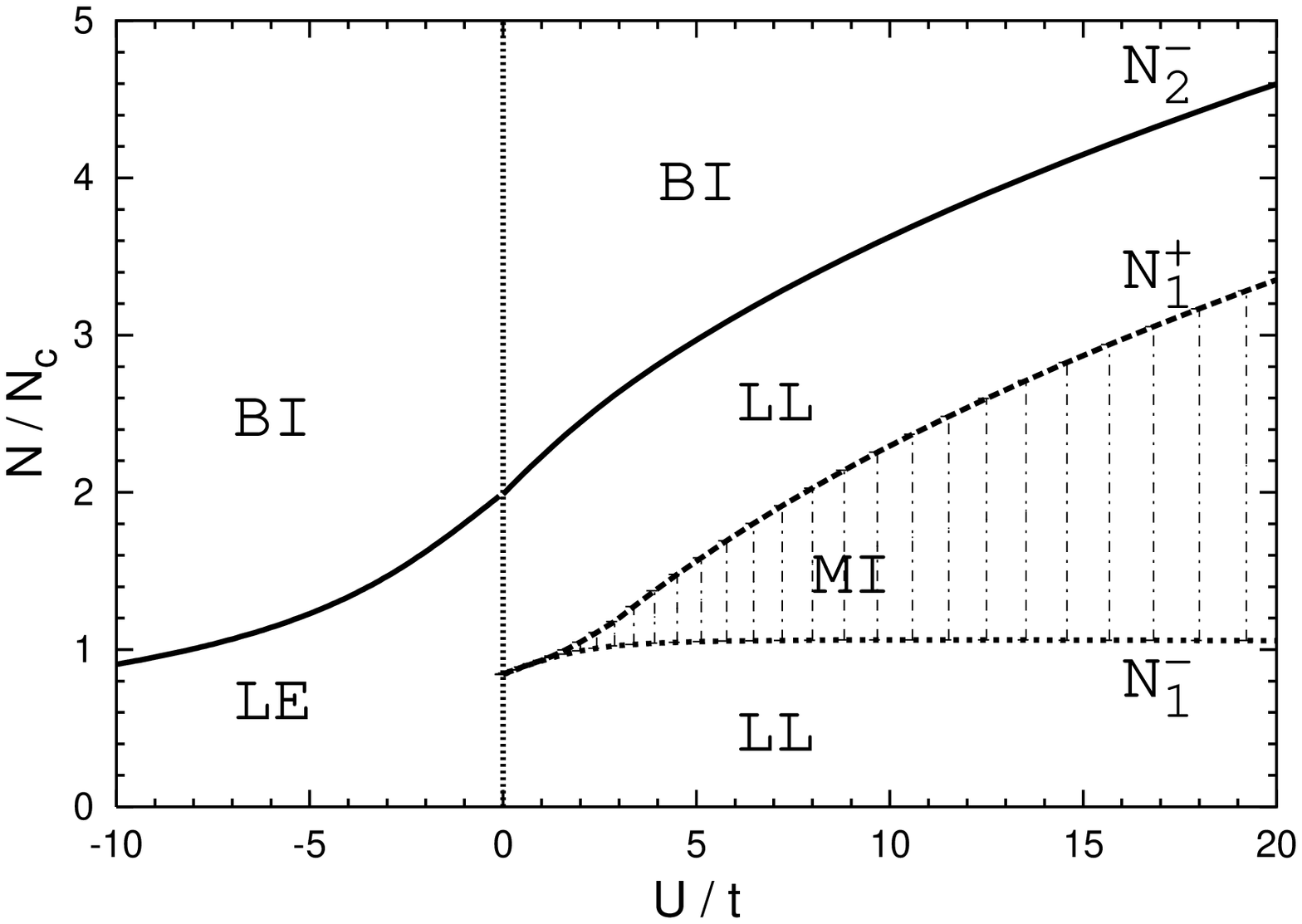,height=8.0cm,angle=-90}
\vspace{.2cm}
\begin{caption}
{Phase diagram of Fermions in a confined 1D optical lattice
vs. $U/t$ and number of atoms.  The phases are denoted by
the core state. For example, for $U>0$ the
BI phase has a band insulator core with
$\rho=2$) surrounded by a LL layer with $1<\rho<2$ again surrounded by a
MI layer with $\rho=1$ and finally a surface LL layer with $0\le \rho<1$.
The curves separate the phases where the MI appear ($N_1$), vanish
($N_1^+$) and a BI ($N_2^-$) appear in the center (see text). 
For $U\le0$ the LE liquid has a BI core when $N>N_2^-$.
}
\end{caption}
\end{center}
\label{f1}
\end{figure}
\vspace{-.2cm}

\subsection{1D Bosons}

Bosons do not experience Pauli blocking and therefore
have a rather different phase diagram as fermions even for the same
lattice, particle mass and interactions.
The Bose-Hubbard \cite{Fisher} is given by:
\bea
  H= -t\sum_{\av{ij}} \hat{b}_{i}^\dagger \hat{b}_{j} \,+\,
  \frac{1}{2}U\sum_i \hat{n}_{i} (\hat{n}_{i}-1) \,,
\eea
where $\hat{b}_i$ is the Bose creation operator and
$\hat{n}_{i}=\hat{b}_{i}^\dagger \hat{b}_{i}$ the density.

Superfluid transitions to
MI solid  phases appear at integer filling, $n=1,2,3,...$,
above a critical coupling \cite{Scalettar,Monien} $U_1=3.84t$, $U_2=6.2t$,   
and $U_n=2.2nt$ for $n\gg1$.
Just above a critical coupling a Mott gap opens in the chemical potential
compatible with a Kosterlitz-Thouless transition: 
$\Delta\mu_n\sim \exp(-const/\sqrt{U-U_n})$. When $U\gg U_n$ we have
$\Delta\mu_n=U-2zt$, whereas for $U\le U_n$ there is no gap and
$\Delta\mu_n=0$.
A good approximation to numerical results \cite{Scalettar}
for $U\gg t$ is the simple parametrization
\bea \label{eB}
 \mu =  nU+(U-\Delta\mu_n)\sin^2[(\rho-n)\pi/2] -zt \,.
\eea
Here, $n=Int[\rho]$ is the integer value of the density. The EoS 
is constructed to have the correct chemical potential gap $\Delta\mu_n$ 
built in as well as the correct critical coupling $U_n$. Therefore the EoS and resulting
phase diagrams are expected to be more accurate than, e.g., calculations within the 
mean field approximation.
The compressibility $d\rho/d\mu$ vanishes in the MI but diverges for densities
approaching the MI as in \cite{Scalettar}.
For $U\gg t$ it has the same EoS as that for fermions, namely that of a 
non-interacting gas of a single fermion species. The strong on-site
repulsion acts as Pauli-blocking similar to the case of 
strongly correlated nuclear liquid at high densities.
The EoS can give the 1D Bose phase diagram at least
qualitatively and generalizes the one shown in \cite{Kollath}.

The critical radii for the MI transitions can now be calculated in the
same way as for fermions by inserting Eq. (\ref{eB}) in
Eq. (\ref{mu}). In the strong coupling limit ($t\to 0$) the Boson 
EoS of Eq. (\ref{eB}) becomes exact and we find
\bea \label{Rn+}
   R_{n}^+ = \sqrt{nU/V_2} \,,
\eea
and
\bea \label{Rn-}
   R_n^- = \sqrt{[nU-\Delta\mu_n]/V_2} \,.
\eea
The corresponding critical number of particles 
$N_n^\pm=(2/d)\int^{R_n^\pm}_0\rho(r)dr$
are shown in Fig. 2. When $U\gg t$ and $n\gg1$
\bea
   N_{n}^+ =N_{n+1}^-=\frac{4n}{3d}\sqrt{\frac{nU}{V_2}} \,.
\eea
Note, that $R_1^-=R_c$ and  $N_1^-=N_c$ for $U\gg t$.

\vspace{-0.5cm}
\begin{figure}
\begin{center}
\psfig{file=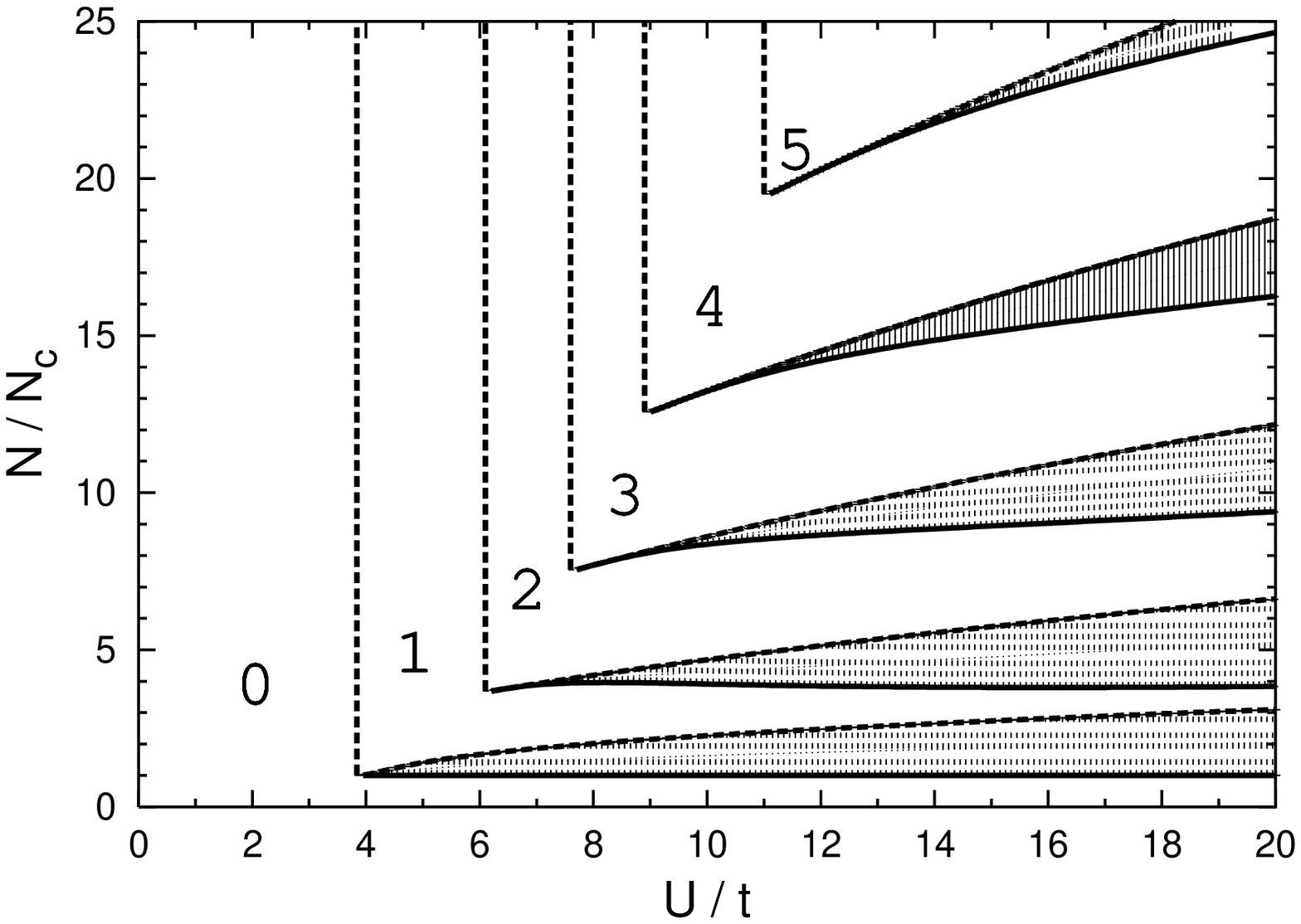,height=8.0cm,angle=-90}
\vspace{.2cm}
\begin{caption}
{Phase diagram for 1D bosons with EoS of Eq. (\ref{eB}). 
The parameter regions numbered $n=0,1,2,3,..$ 
form $n$ MI phases of density $\rho=(0),1,2,..,n$ 
separated by $n+1$ metal regions between core and surface. 
In the hatched areas the density has a MI
core $\rho(0)=1,2,3,..$, from below and up.
}
\end{caption}
\end{center}
\label{f2}
\end{figure}
\vspace{-.2cm}

\subsection{Higher dimensions}

In higher dimensions $D=2,3$ the radii and critical fillings may be
estimated from the scale of the energy and chemical potential.  At
densities up to unit filling the chemical potential varies monotonously 
from $-2zt$ up to $0$ or $+2zt$ in the weak and strong coupling limits
respectively. The chemical potential thus varies on a scale
$\sim t$ and therefore the size of the system remains of order
$R_{1}^-\sim\sqrt{2zt/V_2}$. Above unit filling the chemical potential
gaps are of order $\sim U$ when $U\gg t$ and therefore the scale for
the critical radii are unchanged
\bea \label{Run}
  R_{n+1}^-= R_{n}^+ = \sqrt{n}R_u \,\quad , \, U\gg t \,,
\eea
where $R_u=\sqrt{U/V_2}$ is the critical radius for strong couplings
(see Eqs. (\ref{Rn+}-\ref{Rn-})). 
The corresponding critical number of
particles is
\bea
   N_{n+1}^-=N_n^+=
   \frac{1}{d^D}\int\rho(r)d^D{\bf r} = 
   N_u \sum_{m=0}^n m^{D/2}\, , 
\eea
where $N_u=\kappa_D(R_u/d)^D$, with geometrical factor
$\kappa_D=2,\pi,4\pi/3$ in $D=1,2,3$ dimensions respectively.

 The phase diagram may have additional
phases as in the case of the 2D Hubbard model, which is believed to
describe high temperature superconductivity.
With extended interactions and hopping more MI phases at fractional
filling \cite{FQHE} will appear which may separate into supersolid
phases \cite{Scarola}. Mixing bosons and fermions leads to a 
number of new phases \cite{Lukin}.
Spin imbalance inhibits pairing (as isospin asymmetry in atomic nuclei)
but may also lead to novel pairing
phases \cite{Zwierlein,Hulet}. 

\subsection{High temperature superconductivity}

If high Tc is described by the 2D Hubbard model
\cite{Scalettar,Hofstetter}, then we can expect that the MI is
replaced by an AFM phase around $\rho=1$ surrounded by two
dSC phases at densities lower (underdoped) and
higher (overdoped). According to \cite{Capone} phases with AFM and dSC
order are mixed when $U\la8t$ but coexist when $U\ga8t$ with a first
order transition between two densities.
A critical point is predicted where a mixed AFM+dSC phase terminate
and is replaced by a first order phase transition. Here the
AFM and dSC phases are separated and coexist with
a density discontinuity. In \cite{Aichhorn}, however, the pure dSC 
undergo a first order transition to a mixed AFM+dSC phase.
The densities at which these transitions take place are around
doping $x=1-\rho\simeq \pm 0.1$.
If nearest-neighbor interactions or next-nearest-neighbor hopping are 
included, the particle-hole symmetry is broken and the phase diagram becomes
asymmetric around half filling ($\rho=1$). 

 The corresponding phases in
confined optical lattices will therefore be much more complex than
the 1D shown in Fig. 1. For example, the LL phases will be
replaced by metal, dSC and AFM phases and mixed phases. 
According to \cite{Capone} there is a also a mixed dSC+AFM phase between
the pure dSC and AFM phases when $U\la8t$ whereas above the dSC and AFM
is separated by a first order transition line.
The associated density discontinuity is in stark contrast to the 
constant density MI phase.

Whereas the details of the phases around the MI and AFM shells are not
well known, the surrounding shells of
dSC phases are again surrounded by two metal phases
extending to the surface and the BI core respectively. It would
most interesting to study this phase diagram for the 2D Hubbard model
on the lattice and to find these phases and
possible other phases such as spin glass, strange metal, striped,
checker board, etc., and see how they vary with coupling $U/t$, density
and temperature.

\section{Observables}

The phases described above can be observed by time-of-flight
expansion, interference, noise correlations, molecule formation,
\cite{Greiner,Stoferle,Gerbier,Thalhammer}, collective modes
\cite{Kinast,Hu,mode}, rotation \cite{Zwierlein}, etc.  We discuss
density distributions and radii, collective modes and rotation below.

\subsection{MI shell radii}

The density distribution has recently been measured 
\cite{Folling,Campbell} for 3-D bosons and
the MI plateaus observed directly up to $n=5$. The radii of the MI shells were
measured for different atomic numbers for a large coupling  \cite{Folling}.

In the strong coupling regime the radii of the MI shells $n=1,2,...$ are
(see Eq. (\ref{Run}) and Ref. \cite{DeMarco})
\bea
   R_{n}^2= R_1^2-(n-1) R_u^2 \,.
\eea
Note that the radii $R_n$ are continous functions of filling whereas
the critical radii $R_n^\pm$ are related to a critical filling $N_n^\pm$.

The corresponding particle number in D dimensions is
\bea
  N = N_u \sum_{n=0} \left( \frac{R_1^2}{R_u^2}-(n-1) \right)^{D/2} \,.
\eea
For bosons the
upper limit in the sum is given by
$n-1\le (R_1/R_u)^2$ such that $R_{n}\ge0$. For fermions also $n\le2$.

 The radii and particle number can now be related universally,
i.e. independent of $U$, $t$, $V_2$ and $d$ separately
as long as we are in the strong coupling
regime, when scaled by $R_u$ and $N_u$ respectively. The curves are
shown in Fig. 3 for the 3-D case at zero temperature. 
In the recent experiment of \cite{Folling} a 3D optical lattice is filled
with $^{87}Rb$ atoms in the strong coupling regime such that
$R_u/d=\sqrt{U/V_2d^2}\simeq26$. The measured radii vs. filling is also 
shown for comparison in Fig. 3 and are in good agreement with theory.

\vspace{-0.5cm}
\begin{figure}
\begin{center}
\psfig{file=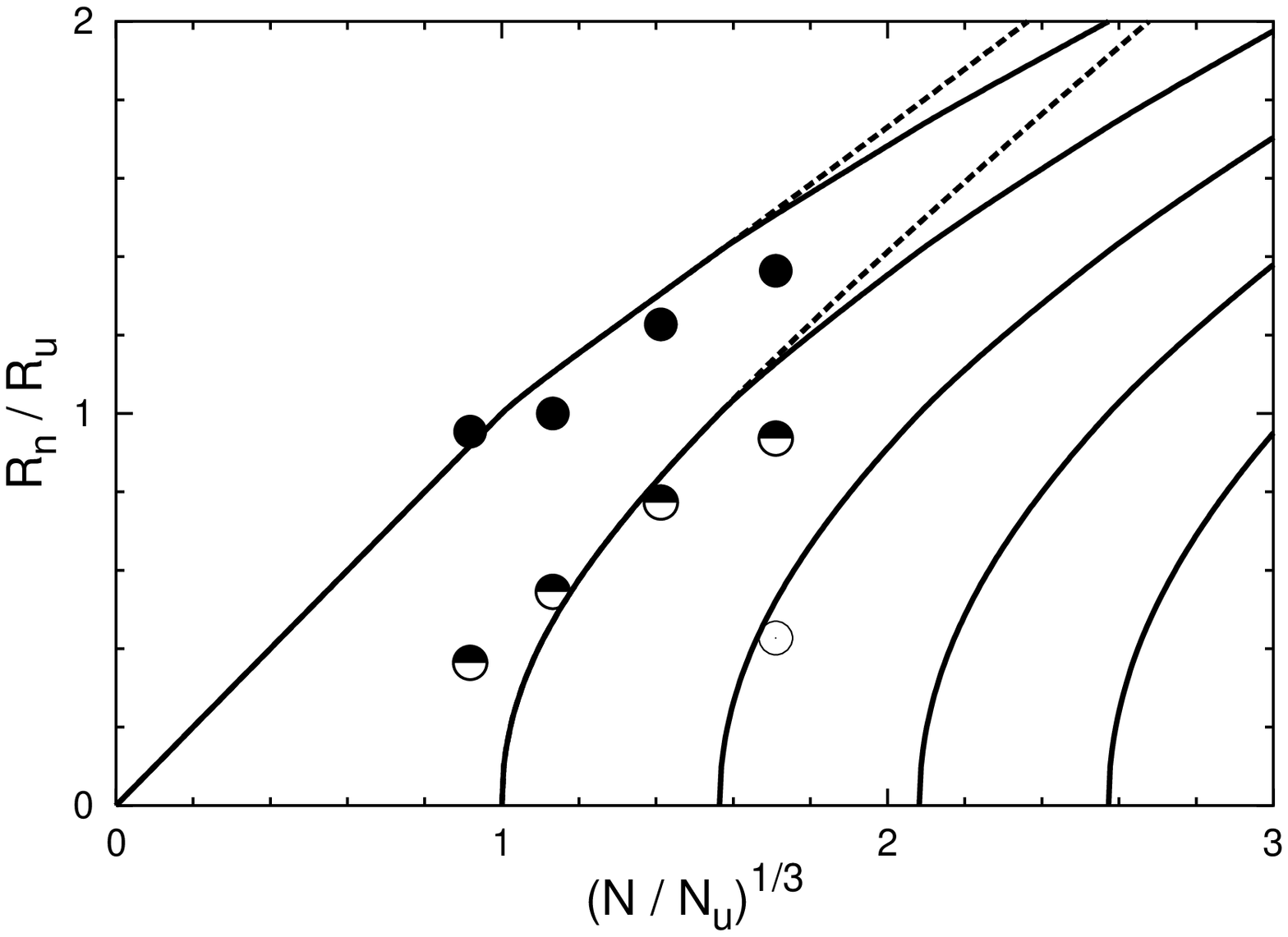,height=8.0cm,angle=-90}
\vspace{.2cm}
\begin{caption}
{Radii $R_n$ for MI shells vs. number of particles for a 3D optical lattice.
Full curves from left to right show the $n=1,2,..5$ shell radii 
for bosons. The dashed curves
show the MI $n=1$ and BI $n=2$ for fermions. 
The solid, half filled and hollow data points
are the first three bose MI radii measured in Ref. [5].
%\cite{Folling}.
}
\end{caption}
\end{center}
\label{f3}
\end{figure}
\vspace{-.2cm}

\subsection{Collective modes}

The collective modes has
proven very useful observable in traps \cite{Kinast,Hu,mode} and we
will therefore calculate the modes for optical lattices.  

The equations of
motion are in hydrodynamics and for a superfluid given by the equation
of continuity and the Euler equation.  Linearizing around equilibrium
these equations lead to \cite{PS,mode}
\bea \label{EOM}
   -m\omega^2 {\bf v} = \frac{1}{\rho}{\bf \nabla}
  \left[\rho^2 \frac{d\mu}{d\rho}{\bf \nabla\cdot v}\right] 
     +\nabla({\bf f\cdot v})-{\bf f}(\nabla{\bf \cdot v}) \,,
\eea
for the flow velocity ${\bf v}$. Here, the force from the 
harmonic trap potential is ${\bf f}=-\nabla(V_2r^2)=-m\omega_0^2 {\bf r}$,
where we have introduced the trap frequency
$\omega_0$ through $V_2=m\omega_0^2/2$.

The collective modes can be calculated analytically from
Eq. (\ref{EOM}) when the EoS is a simple polytrope:
$\mu(\rho)-\mu(0)\propto\rho^\gamma$.  At low densities the EoS
optical lattices is on such a form for fermions with $\gamma=2/D$. In
1D $\gamma=2$ for both fermions and bosons according to Eq. (\ref{eF})
and (\ref{eB}).  At higher densities the effective polytrope
$\bar{\gamma} \equiv [\rho/(\mu-\mu(0))]d\mu/d\rho$ taken at central
density $\rho(0)=\rho_c$ gave rather accurate collective modes in atomic
traps \cite{Hu,mode}. For 1D fermions in an optical lattice
we find from Eq. (\ref{eF})
\bea \label{gamma}
  \bar{\gamma} 
%\equiv \left( \frac{\rho}{\mu-\mu(0)}\frac{d\mu}{d\rho}\right)_{\rho=\rho_c} 
    = \frac{\pi\rho_c}{\beta} \cot(\pi\rho_c/2\beta) \,,
\eea
for $\rho_c<1$. In the weakly interacting and the attractive cases $U\le0$
with EoS given by Eq. (\ref{LE}), we also obtain Eq. (\ref{gamma})
with $\beta=2$. 

The collective modes of fermions and bosons are the same in the strong
coupling limit since bosons with EoS of
Eq. (\ref{eB}) also has the effective polytrope of Eq. (\ref{gamma})
with $\beta=1$.
In contrast a dilute bose EoS ($U\ll U_1$) has
$\bar{\gamma}=1$ (see Fig. 3). We can expect a continuous change of
$\bar{\gamma}=1$ between these two limits.  When $\rho_c\to1$ and
$U\gg t$ both bosons and fermions have $\bar{\gamma}\to0$ whereas at
$\rho_c=1$ a MI forms in the center making the system incompressible
(corresponding to $\bar{\gamma}\to\infty$).

As in 3D \cite{mode} the collective modes can be calculated in 1D
generally for any polytropic EoS. Eq. (\ref{EOM}) reduces to
\bea \label{v}
 \frac{1}{2}(1-x^2)\frac{d^2v}{dx^2} -(\gamma+1)x\frac{dv}{dx}+
  (\frac{\omega^2}{\omega_0^2}-1)v =0 \,,
\eea
where $x=r/R$. The eigenfunctions are
$v \propto F(-n,n+2\gamma-1,\gamma+1,2x-1)$,
where $F$ is the hypergeometrical function. The corresponding
1D eigenvalues for multipoles with $n=0,1,2,...$, nodes are
\bea \label{1D}
  \frac{\omega^2}{\omega_0^2} = 1+n[\gamma+1+(n-1)/2] \,.
\eea 

The dipole mode ($n=0$) has constant flow velocity in all dimensions and by
insertion in Eq. (\ref{EOM}) has frequency $\omega/\omega_0=1$. It
is independent of dimension, EoS ($\gamma$) and central density because the
dipole mode corresponds to the whole cloud sloshing as a rigid body in
the trap and thus only depends on the trap geometry in one direction
(as long as $\rho_c<1$).
The breathing or monopole mode frequency has
flow velocity ${\bf v}\propto {\bf r}$ and
by insertion in Eq. (\ref{EOM})
the monopole frequency is found to be
$\omega^2/\omega_0^2 = 2+\gamma D$, generally 
in $D$-dimensions.

\vspace{-0.5cm}
\begin{figure}
\begin{center}
\psfig{file=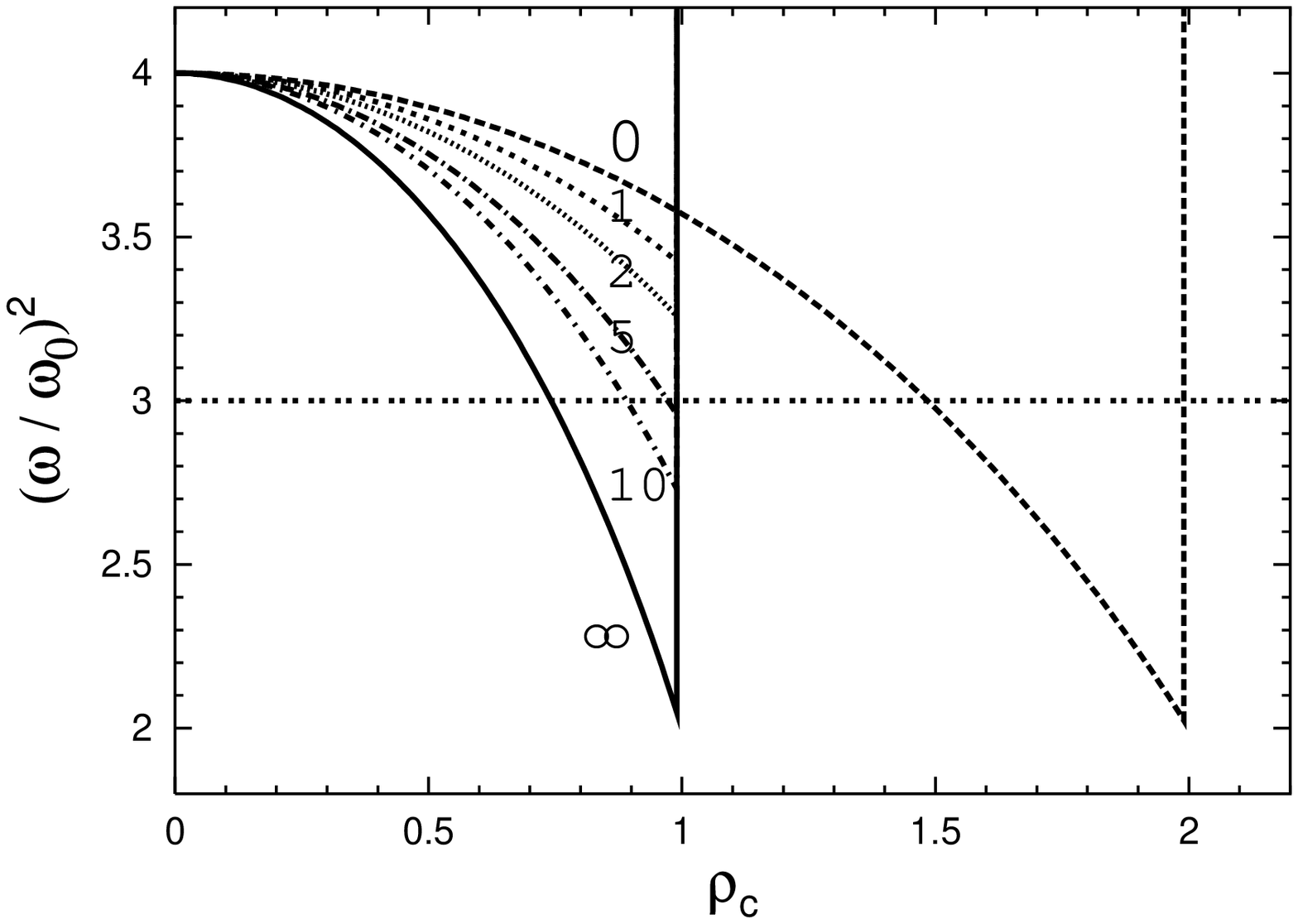,height=8.0cm,angle=-90}
\vspace{.2cm}
\begin{caption}
{1D breathing mode frequency $\omega^2/\omega_0^2=2+\bar{\gamma}$
vs. central density. The modes for fermions are labelled with
$U/t=0,1,2,5,10,\infty$. 
Full curve gives the strong coupling limit $U\gg U_1$ for both fermions and
bosons. The weak coupling limit $U\simeq0$ 
is shown by dashed curve for fermions
and dotted curve for bosons where
$\omega^2/\omega_0^2=3$.
}
\end{caption}
\end{center}
\label{f4}
\end{figure}
\vspace{-.2cm}

In Fig. 3 we show the 1D breathing mode frequency ($n=1$)
$\omega^2/\omega_0^2=2+\bar{\gamma}$ with the effective polytrope
$\bar{\gamma}$ given by Eq. (\ref{gamma}). It decreases with
increasing density until the MI appears in the core which is
incompressible. Likewise we can expect that the collective modes will
reveal the AFM state for the 2D Hubbard model. The dSC
may only expected to affect the EoS and collective modes
weakly. Rotational vortices \cite{Zwierlein} or RF spectroscopy
\cite{Chin} may therefore be better signals for the dSC phases.

\subsection{Rotation}

By rotating the atoms in the traps the BEC and BCS superfluids are
revealed by vortex formation \cite{Zwierlein}. Similarly, for confined
2D optical lattices the layers of BEC or dSC superfluids with bosons and
fermions respectively may be identified by vortices when rotated. The
MI phase is solid and the atoms in it cannot be rotated unless a
supersolid phase exists or the lattice itself is rotated.

The possible density discontinuity \cite{Capone,Aichhorn} leads to a
non-analytic moment of inertia. Due to centrifugal forces the critical
radii and fillings now also depends on rotation frequency and
therefore the density discontinuity can made to occur right in the
center by varying either the particle number or the angular frequency.
Such core phase transitions have been considered for neutron stars
resulting in a characteristic behavior of neutron star spin down
frequencies with time \cite{NS}. 

Solids in confined optical lattices
will prefer ellipsoidal deformation at fast rotation and varying the
angular velocity may therefore create glitches (MI quakes) as observed
for neutron stars.

\section{Summary}

In summary, optical lattices provide a clean system where 1, 2 and 3D
Hubbard and Bose-Hubbard models can be realized.  The phase
transitions between MI, LL, LE, superfluids and metals were calculated
in 1D and in 2+3D for the strong coupling regime, where comparison was
made to recent data on radii and fillings for bose atoms in optical lattices.
Collective modes were calculated
and found to be sensitive to the underlying EoS and phases.  2D
optical lattices with fermions will therefore give direct insight in
the high temperature superconductivity and the dSC+AFM phases vs.
interaction strength, doping and temperature.

\vspace{-.4cm}
%\section*{References}

\end{document}